\documentclass[11pt,a4paper]{JHEP3}

\newcommand{\be}{\begin{equation}}
\newcommand{\ee}{\end{equation}}
\newcommand{\ch}{{\cal H}}
\newcommand{\half}{{1 \over 2}}
\newcommand{\bea}{\begin{eqnarray}}
\newcommand{\eea}{\end{eqnarray}}
\newcommand{\co}{{\cal O}}
\newcommand{\ci}{{\cal I}}
\def\cn{{\cal N}}
\def\ct{{\cal T}}
\def\hP{\hat{\Psi}}
\def\O{\Omega}
\def\({\mbox{$(\!($}}
\def\){\mbox{$)\!)$}}
\def\Tr{{\rm Tr}\, }
\def\X{\Xi}

\title{Ghost Sector of Vacuum String Field Theory and the Projection Equation}
\author{Robertus Potting\\ \'Area Departamental de F\'\i sica / CENTRA\\
 FCT, Universidade do Algarve\\  Campus de Gambelas, 8000-117 Faro, Portugal\\
 E-mail: \email{rpotting@ualg.pt}}
\author{Joris Raeymaekers\\Department of Theoretical Physics\\
Tata Institute of Fundamental Research\\
Homi Bhabha Road,
Mumbai 400005, INDIA\\E-mail: \email{joris@theory.tifr.res.in}}
\preprint{hep-th/0204172}
\abstract{We study the ghost sector of vacuum string field 
theory where the BRST operator $Q$ is given by the midpoint insertion
proposed by Gaiotto, Rastelli, Sen and Zwiebach.
We introduce a convenient basis of half-string modes in terms of which 
$Q$ takes a particularly simple form.
We show that there exists a field redefinition
 which reduces the ghost sector field equation to a
pure projection equation for string fields
satisfying the constraint that the ghost number
is equally divided over the left- and right halves of the string.
When this constraint is imposed, vacuum string field theory
can be reformulated as a $U(\infty)$ cubic matrix model.
Ghost sector solutions can be constructed from 
projection operators on half-string Hilbert space just as in the
matter sector. We construct the ghost sector equivalent 
of various well-known matter sector projectors such as the sliver,
butterfly and nothing states.}

\begin{document}

\section{Introduction}

Witten's open string field theory (OSFT) \cite{Witten}
has proven succesful
in desribing the process of tachyon condensation on unstable 
brane configurations. The solution corresponding to the tachyon vacuum
has been approximated to a high degree of accuracy using level truncation
methods \cite{tachcond}. However, an explicit analytic solution is still 
lacking.
Such a solution would be desirable in order to describe the fluctuations
 around the tachyon vacuum, where interesting new physics is 
expected to emerge, possibly including closed strings \cite{closedstr}.
Vacuum string field theory (VSFT) was proposed as an `educated guess'  
for the string field theory around the tachyonic vacuum. Various requirements
for such a theory 
led the authors of \cite{RSZ1} to propose an action of the same cubic
form as in OSFT but with a different BRST-operator $Q$ depending only on the 
ghost modes. This makes it natural to look for solutions $\Psi$ in which the 
matter and ghost parts decouple:
\be
\Psi = \Psi^m \otimes \Psi^{gh}.
\label{ansatz}
\ee
The equation of motion for the matter part then becomes the projection equation
\be
\Psi^m =  \Psi^m \star^m  \Psi^m
\label{projeq}
\ee
while the ghost part has to satisfy
\be
Q \Psi^{gh} =  \Psi^{gh} \star^{gh}  \Psi^{gh}.
\label{eqgh}
\ee

Various checks of VSFT can be made without precise knowledge of $Q$ or 
the ghost part $\Psi^{gh}$. The matter part
of the D-25 brane solution is given by the `sliver state' \cite{KP}
and lower dimensional
D-branes correspond to generalizations thereof. These solutions reproduce
the correct ratios of D-brane tensions \cite{RSZ2, oku}. 
Multi-D-brane solutions have also
been constructed \cite{RSZ3}.

A convenient formulation for finding solutions to the projection equation
(\ref{projeq}) is given in the split-string formalism \cite{bordes, GT1, GT2}. 
Here, it is possible to represent
string fields as operators on an auxiliary `half-string Hilbert space' 
$\ch^\half$
in such a way that the star product becomes the multiplication of operators.
Numerous solutions to (\ref{projeq}) can then be constructed in the form 
of projection
operators on $\ch^\half$. This construction generalizes a familiar technique
from noncommutative field theories \cite{GMS}.
Another approach to constructing solutions
to (\ref{projeq}) uses surface states in conformal field
theory \cite{RSZ4, RSZ5, GRSZ2}.

In \cite{GRSZ1}, Gaiotto, Rastelli, Sen and Zwiebach (GRSZ) presented
 convincing  arguments that point to 
a specific choice for the BRST operator $Q$ in VSFT. 
It is given by a pure midpoint insertion
\bea
Q &=&  c_0 - (c_2 + c_{-2}) + (c_4 + c_{-4}) - \ldots 
\nonumber\\
&=& {1 \over 2} \left( c^+(\pi/2) + c^-(\pi/2) \right).
\label{Qmid}
\eea 
It was also argued that the overall multiplication constant
in front of the action should diverge in order to have finite energy
solutions.

Finding solutions to the ghost sector equation (\ref{eqgh}) of VSFT seems
at first sight more difficult than in the matter sector because
of the presence of $Q$ and the fact that $\star^{gh}$ contains a midpoint  
insertion.
However, solutions were found in \cite{GRSZ1} which correspond to 
projectors in an auxiliary twisted
ghost system.  
In this paper we further explore the relation between 
solutions to the ghost sector equation (\ref{eqgh}) 
with $Q$ given by the midpoint insertion (\ref{Qmid}) and  projection
operators. Our approach uses oscillator methods \cite{GJ1, GJ2} 
and the bosonized
form of the ghosts. After transforming to
suitably chosen half-string variables, we show that there exists
a field redefinition which takes the ghost equation of motion into a
pure projection equation of the form
\be
\Psi' = \Psi' \star' \Psi'.
\label{transfeom}
\ee  
provided that $\Psi'$ satisfies the constraint that the ghost number
should be equally divided over the left- and right halves of the string.
The star product $\star'$ in (\ref{transfeom}) is different from
the original ghost sector product $\star^{gh}$ in that it no longer includes
midpoint insertions: it is given by a pure delta-function overlap just
as in the matter sector. This allows us to construct solutions from projection
operators 
on half string Hilbert space as in the matter sector. In this way several 
well-known matter sector solutions have a direct counterpart in the ghost
sector. As an illustration, we construct the ghost sector equivalent 
of various well-known matter sector projectors such as the sliver,
butterfly and nothing states.

This paper is organized as follows. Section \ref{matter} deals with the 
transformation to half-string modes which we illustrate in the matter 
sector. We are careful to choose a half-string basis which gives well 
defined expressions for midpoint insertions. We discuss the 
Bogoliubov transformation relating full- and half-string
oscillator modes and derive the relation
between full-string and half-string vacua. We also find expressions for
the Neumann matrices in the half-string basis.  
In section \ref{examples}, we 
discuss several solutions to the matter sector projection equation 
which will turn out to have a direct counterpart in the ghost sector.
In section \ref{ghost}, we turn to the transformation to half-string variables 
in the ghost sector. We derive the bosonized version of the VSFT BRST-operator
which takes a simple form in terms of our half-string variables. 
In section \ref{redef}, we discuss the field redefinition which
turns the ghost sector equation of motion into a pure projection equation 
for string fields satisfying the constraint that the ghost number is 
equally divided into the left and right-parts. 
When this constraint is imposed, vacuum string field theory
can be reformulated as a $U(\infty)$ cubic matrix model.
 We  
 comment on the construction of solutions and
give several examples. We end with a discussion of some open problems.  

\section{Vacuum string field theory: matter sector}\label{matter}

\subsection{Mode expansions}
We start this section by giving some conventions regarding mode expansions.
The string is parametrized by $\sigma \in [0, \pi]$ and mode expansions
are obtained by expanding in orthonormal basis functions $\psi_n (\sigma)$ 
of $L_2 [0, \pi]$
obeying Neumann boundary conditions at the endpoints:
\be \psi_0 = 1 ; \qquad \psi_n = \sqrt{2} \cos n \sigma \label{basis}.\ee
One obtains position and momentum modes by expanding (we will suppress
Lorentz indices in the rest of the paper):
\bea
X (\sigma) &=& \sum_{n = 0}^{ \infty}
x_n \psi_n(\sigma) = x_0 + \sqrt{2} \sum_{n = 1}^{ \infty}x_n \cos n \sigma\nonumber\\
\pi P (\sigma) &=& \sum_{n = 0}^{ \infty}
p_n \psi_n(\sigma) = p_0 + \sqrt{2} \sum_{n = 1}^{ \infty}p_n \cos n \sigma.\nonumber
\eea
We represent the modes $\{x_n\}_n$ as a column vector
$|x)$ and use the notation $(x|y) \equiv  \sum_{n=0}^\infty x_n y_n$. 
Creation-annihilation operators are defined by:
\bea
|x) = {i \over \sqrt{2}} E [ |a) - |a^\dagger) ]&\qquad&
|p) = {1 \over \sqrt{2}} E^{-1} [ |a) + |a^\dagger) ]\nonumber\\
|a) = {1 \over \sqrt{2}} [ E |p) - i E^{-1} |x)]&\qquad&
|a^\dagger) = {1 \over \sqrt{2}} [ E |p) + i E^{-1} |x)]
\label{fullcreation}
\eea
with $E$ the matrix with coefficients
$$
E^{-1}_{mn} = \sqrt{{n}} \delta_{mn} + \sqrt{2} \delta_{m0} \delta_{n0}
$$
The canonical commutation relations are:
$$
[a_m, a_n^\dagger]= \delta_{mn}
$$
As usual, one builds up a Fock space by acting with the creation operators
on a vacuum state $|\O\rangle$ annihilated by
 all $a_n$ (including $a_0$) and normalized to
$\langle\O|\O\rangle =1$. Sometimes it is useful to work with the 
translationally invariant vacuum
$|0\rangle$  satisfying $p_0 |0\rangle = a_n |0\rangle = 0,\ n>0$.
Position basis states can be expressed in terms of oscillators
as
\be
|x\rangle \equiv |\{x_n\}_n \rangle = N^X \exp \left[
 - \half (x | E^{ -2}| x) - 
\sqrt{2} i (a^\dagger|
E^{-1}|x) + \half (a^\dagger|a^\dagger)\right] |\O\rangle
\ee
with $N^X$ an (infinite) normalization constant:
$$
N^X = \left({2 \over \pi}\right)^{26/4} \prod_{n=1}^\infty \left({n \over \pi}
\right)^{26/4}.
$$
\subsection{Split-string formalism}
In this section we set up the transition to the split-string formalism.
This transition is usually \cite{bordes, GT1} obtained by 
splitting $X(\sigma)$ into
 left and right parts
\bea
X^L(\sigma) &=& X(\sigma) \qquad 0 \leq \sigma \leq \pi/2\\
X^R(\sigma) &=& X(\pi - \sigma) \qquad 0 \leq \sigma \leq \pi/2
\eea
and performing mode expansions of these functions on the
interval $[0,\pi/2]$, where one chooses some boundary condition in the
midpoint $\sigma = \pi/2$. However, this leaves some ambiguities
in the treatment of the midpoint: for instance, it is not clear
how to express $X(\pi/2)$ in terms of half-string modes.
Our setup is slightly different   in the sense 
that we will treat the transition to half-string modes as a change of 
basis {\em in the full interval $[0,\pi]$}. 
The change of basis is chosen
so as to diagonalize the projection operators on the left and right
 halves of the string. 
Various choices of orthonormal half-string bases are possible and
the expansion of a string configuration $X(\sigma)$ in any such  basis 
will converge
with respect to the $L_2$ norm. However, even if $X(\sigma)$ is continuous
in $\sigma = \pi/2$,
pointwise convergence of the
expansion in $\sigma = \pi/2$
is not guaranteed.  Due to the importance of the midpoint in string 
field theory (especially in the
ghost sector where midpoint insertions enter explicitly), 
we will impose pointwise convergence at the midpoint
as an additional criterion for the half-string basis we will use. 
In a basis satisfying this criterion, $X(\pi/2)$
has a well-defined expansion in terms of half-string modes.
\subsubsection{Half-string projectors}
Instead of working with the basis (\ref{basis}), one can also set up an 
expansion in a different  orthonormal basis 
in which the projection operators $P^L, \ P^R$ on the 
left- and right parts 
of the string become diagonal. The coefficients in such an
 expansion are referred to as 
half-string modes. 
The matrix elements of $P^L (\sigma, \sigma ') = \theta (\pi/2 - \sigma) \delta(\sigma - \sigma ')$
with respect to the basis (\ref{basis}) are easily obtained.
Splitting the component indices in even and odd ones 
(denoted by superscripts $^e$ and $^o$ respectively) 
and writing any matrix $M$ as $M = \left( \begin{array}{cc} M^{ee}
& M^{eo} \\ M^{oe} & M^{oo} \end{array} \right)$, one finds
$$ P^L= \half \left( \begin{array}{cc} 1^{ee} & A \\ A^T & 1^{oo}
 \end{array} \right).$$
Here we have defined a matrix $A$ whose rows are labeled by 
even indices and whose columns are labeled by odd ones\footnote{The 
matrix $X$ defined in \cite{GT1} is related to our $A$ by 
$X = \left( \begin{array}{cc} 
0^{ee} & A \\ A^T & 0^{oo}
 \end{array} \right)$.}
\bea
A_{2m, 2n+1} &=& { 4 (2n+1)  (-1)^{m+ n} \over
((2n + 1)^2 - (2m)^2) \pi }\nonumber\\
 A_{0,2n+1} &=&{2\sqrt{2} (-1)^n \over  (2n + 1) \pi}
\label{Adef}
\eea
The matrix $A$ satisfies $A A^T= 1^{ee},\ A^T A =1^{oo}$, 
hence $(P^L)^2 = P^L$.\\ 
\ \\
Similarly, for the projection operator on the right half of the string
$P^R (\sigma, \sigma ') = \theta (\sigma - \pi/2) \delta(\sigma - \sigma ')$, one finds
the component form
$$ P^R = 1 - P^L =\half \left( \begin{array}{cc} 1^{ee} & -A \\ -A^T & 1^{oo}
 \end{array} \right).$$

\subsubsection{Half-string basis}\label{half}
The Hilbert space $L_2 [0, \pi]$
decomposes into the direct sum of $P^L$ and 
$P^R$-invariant
subspaces. Our goal is to
 introduce a new orthonormal basis adapted to this decomposition.
Consider the matrix  
$$ \co  = {1 \over \sqrt{2}} 
\left( \begin{array}{cc} 1^{ee} & A \\ 1^{ee} & -A \end{array} \right).$$
It is orthogonal, $\co \co^T = \co^T \co = 1$ and diagonalizes $P^L, P^R$:
$$ \co P^L \co^T = \left( \begin{array}{cc} 1^{ee} & 0^{ee} \\ 0^{ee} & 0^{ee}
 \end{array} 
\right), \qquad
\co P^R \co^T = \left( \begin{array}{cc} 0^{ee} & 0^{ee} \\ 0^{ee} & 1^{ee}
 \end{array} 
\right).$$
The matrix $\co$ defines a transformation to a new orthonormal basis 
$ \psi^L(\sigma), \ \psi^R(\sigma) $ satisfying $ P^L \psi^L_{2n} 
= \psi^L_{2n},\ P^R \psi^R_{2n} = \psi^R_{2n}$ and  $P^L \psi^R_{2n}
= P^R \psi^L_{2n} = 0$ :
\bea 
\psi^L_{2n}(\sigma) &=& {1 \over \sqrt{2}} [ \psi_{2n}(\sigma) + \sum_m A_{2n,2m+1} 
\psi_{2m+1}(\sigma) ] \\
\psi^R_{2n}(\sigma) &=& {1 \over \sqrt{2}} [ \psi_{2n}(\sigma) - \sum_m A_{2n,2m+1} 
\psi_{2m+1}(\sigma) ] 
\eea
Explicitly, one has
\bea
\psi^L_0(\sigma) = \sqrt{2}\ \theta( \pi /2 - \sigma );&\ &
 \psi^R_0(\sigma) = \sqrt{2}\ \theta( \sigma - \pi/2 )\nonumber\\ 
\psi^L_{2n}(\sigma) = 2 \cos 2n \sigma\ \theta( \pi /2 - \sigma ); &\ &
\psi^R_{2n}(\sigma) = 2 \cos 2n \sigma\ \theta( \sigma - \pi/2 )\ \ \ n \neq 0 \nonumber
\eea
where the step function in $0$ is to be interpreted as
$\theta(0) = \half$.
Hence this choice of half-string basis corresponds to expanding $X^L(\sigma),\ 
X^R(\sigma)$
in basis 
functions on $[0, \pi/2]$  obeying Neumann boundary 
conditions at the 
midpoint.
The mode expansion in the new basis reads
\be
X(\sigma) = \sum_{n=0}^\infty x^L_{2n} \psi^L_{2n}(\sigma) +
\sum_{n=0}^\infty x^R_{2n} \psi^R_{2n}(\sigma) 
\label{halfexp}
\ee
The transformation formulae for the position modes take the form:
\be
|x^L) = {1 \over \sqrt{2} } [ |x^e) + A |x^o) ] \qquad
|x^R) = {1 \over \sqrt{2} } [ |x^e) - A |x^o) ]\nonumber
\label{xtransf}
\ee
The inverse transformations are:
\be
|x^e) = {1 \over \sqrt{2} } [ |x^L) +  |x^R) ]\qquad
|x^o) = {1 \over \sqrt{2} }A^T [ |x^L) - |x^R) ]\nonumber
\label{xtransf_inv}
\ee
This half-string basis satisfies our criterion for pointwise convergence
in the midpoint, indeed one has:
$$
X(\pi/2) = {1 \over \sqrt{2}}(x^L_0 + x^R_0) +  \sum_n (-1)^n 
(x^L_{2n} + x^R_{2n}) = 
x_0 + \sqrt{2} \sum_n (-1)^n x_{2n}.
$$ 
The same procedure can be followed for the decomposition of
the conjugate momentum $P(\sigma)$ into half-string modes. One gets
\bea
|p^L) = {1 \over \sqrt{2} } [ |p^e) + A |p^o) ]&\qquad&
|p^R) = {1 \over \sqrt{2} } [ |p^e) - A |p^o) ]\nonumber\\
|p^e) = {1 \over \sqrt{2} } [ |p^L) +  |p^R) ]&\qquad&
|p^o) = {1 \over \sqrt{2} }A^T [ |p^L) - |p^R) ]
\label{ptransf}
\eea
Due to the orthogonality of the transformation, the canonical
commutation relations are unmodified:
$$
[x^L_{2m}, p^L_{2n}] = i \delta_{mn},\ \ \ [x^R_{2m}, p^R_{2n}] =i \delta_{mn},\ \ \ 
 [x^L_{2m}, p^R_{2n}]=[x^R_{2m}, p^L_{2n}]=0.
$$

Other half-string bases 
can be obtained by making  further orthogonal 
transformations that do not mix the left- and right basis vectors. 
For instance, one could diagonalize the projection operators $P^L,\ P^R$
with an orthogonal matrix $\tilde \co$ defined by
$$
\tilde \co = \left( \begin{array}{cc} A^T & 0^{oe} \\ 0^{eo} &
 A^T \end{array} \right) \co = {1 \over \sqrt{2}} \left( \begin{array}
{cc} A^T & 1^{oo} \\ A^T & -1^{oo} 
\end{array} \right)
$$
The corresponding basis functions are the odd cosines
$$
\tilde \psi^L_{2n+1}(\sigma) =  2 \cos (2n+1) \sigma\ \theta( \pi /2 - \sigma );
 \qquad
\tilde \psi^R_{2n+1}(\sigma) = - 2 \cos (2n+1) \sigma\ \theta( \sigma - \pi/2 ).
$$
This corresponds to expanding $X^L(\sigma),\ 
X^R(\sigma)$
in basis 
functions on $[0, \pi/2]$ obeying Dirichlet
boundary conditions at the midpoint \cite{bordes, GT1, GT2}. 
An expansion in this basis will give zero at $\sigma = \pi/2$,
hence this half-string basis does not
satisfy the criterion of pointwise convergence at the midpoint.
This is often remedied by introducing an extra `midpoint degree of freedom'.
The role of such a degree of freedom is not completely clear since
it doesn't correspond to a basis vector of $L_2[0,\pi]$. We will not
 follow this procedure here but instead use the previous half-string  
expansion (\ref{halfexp}) in the rest of the paper.

As has been noted in the literature \cite{MT, bars, douglas}, the matrix $A$
is not strictly invertible, but has a (nonnormalizable) zero mode:
\begin{equation}
A_{2m+1,0}^T{1\over\sqrt2}+\sum_{n=1}^\infty A_{2m+1,2n}^T (-1)^n=0.
\end{equation}
In the present context, we see from Eq.\ (\ref{xtransf_inv}) that this
means that the half-string mode defined by
\be
x^L_0=-x^R_0={1\over\sqrt2}\qquad\qquad
x^L_{2n}=-x^R_{2n}=(-1)^n\qquad (n>0)
\ee
corresponds to full-string coefficients being equal to zero.
It is not hard to see how this comes about. From the identity
\be
1+2\sum_{n=1}^\infty (-1)^n\cos(2n\sigma)=\pi\delta(\sigma-{\pi\over2})
\ee
it follows that the mode in question corresponds to $X^L(\sigma)=
-X^R(\sigma)\propto\delta(\sigma-\pi/2)$. The two delta functions cancel in
taking the sum (\ref{halfexp}) for $X(\sigma)$.
It is important to keep this over-parametrization in mind in the future.

\subsubsection{String fields as operators}
The formulation of Witten's cubic open string field theory action makes use of
operations involving string fields, the star product $\star$ and the 
integration $\int$, which take a simple form when expressed in terms of
half-string modes. When we regard the string field $\Psi$
 as a functional 
of the half-string modes $\{ x^L_{2n},\ x^R_{2n} \}$ (we will  use the
notation $\Psi [x^L,x^R]$), the $\star$ and $\int$ operations can be written as
\bea
\Psi \star \Phi [x^L, x^R] &=& \int [D y] \Psi[x^L, y] \Phi[y, x^R]\\
\int \Psi &=& \int [D y] \Psi[y, y]
\eea
where $[D y ] \equiv \prod_n d y_{2n}$.
These operations can also be written in operator form as follows.
To any state $|\Psi \rangle$ corresponds 
an operator $\hat \Psi$  through
the correspondence:
\bea 
|\Psi \rangle &=& \int [Dx^L Dx^R]\  \Psi [x^L , x^R] 
|x^L\rangle |x^R\rangle \\
& \Updownarrow & \\
\hat \Psi &=& \int[Dx^L Dx^R] \  \Psi [x^L, x^R] |x^L\rangle 
\langle x^R |
\eea 
Such an operator formally maps a state in the ``right'' half-string Hilbert 
space  to a state in the ``left'' half-string Hilbert
space. Since these spaces can be canonically identified, we will
consider $\hat \Psi$ to be an operator on ``the'' half-string Hilbert 
space which will be denoted by $\ch^\half$.
We will use a subscript $_\half$ to distinguish
states in $\ch^\half$ from their full-string counterparts. Oscillator 
modes in $\ch^\half$ will be denoted by a superscript $^\half$. 
In terms of  operators on $\ch^\half$, star multiplication 
and integration reduce to 
operator multiplication and trace respectively:
\bea
\Phi \star \Psi &\Leftrightarrow& \hat \Phi \hP \\
\int \Psi &\Leftrightarrow& \Tr \hP
\eea
The Hermitean inner product becomes:
\be
\langle \Phi | \Psi \rangle \Leftrightarrow \Tr \hat \Phi^\dagger
\hat \Psi
\ee
In string field theory, one works with string fields for
which the Hermitean inner product $\langle \Phi | \Psi \rangle$ 
becomes equal to the BPZ inner product $\int \Phi \star \Psi$. This
is guaranteed by a reality condition which becomes a Hermiticity condition
in the operator formulation:
$$
\Psi[x^L, x^R] = \Psi^*[x^R,x^L]\Leftrightarrow \hat \Psi = \hat \Psi^\dagger
$$
Summarized, under the assumption (\ref{ansatz}) that matter and ghost parts of 
the string field 
don't mix, the matter part of a solution to the VSFT equations of motion 
 can be represented as an operator satisfying:
\bea
\hat \Psi &=& \hat \Psi^\dagger\\
\hat \Psi &=& \hat \Psi^2
\eea
Numerous solutions can be found by taking  $\hat \Psi$ to be any
Hermitean 
projection operator on half-string space. 
These can, at least in principle, be transformed back to full string
variables using the transformation formulae (\ref{xtransf},\ \ref{ptransf}).
Some relevant examples will be discussed in \ref{examples}.

\subsubsection{Half-string creation and annihilation operators}
In order to get some feeling for the meaning of the transformation to
half-string modes, it is useful to see how it acts on the creation 
and annihilation operators. In analogy with (\ref{fullcreation}), we define
the half-string creation and annihilation operators as 
\bea
|x^L) = {i \over \sqrt{2}} E^{ee} [ |a^L) - |a^{L\dagger}) ]&\ &
|p^L) = {1 \over \sqrt{2}} (E^{ee})^{-1} [ |a^L) + |a^{L\dagger}) ]\nonumber\\
|a^L) = {1 \over \sqrt{2}} [ E^{ee} |p^L) - i (E^{ee})^{-1} |x^L)]&\ &
|a^{L\dagger}) = {1 \over \sqrt{2}} [ E^{ee} |p^L) + i (E^{ee})^{-1} |x^L)]
\nonumber\\
\label{halfcreation}
\eea
and similarly for the right half-string modes.
\\
From (\ref{fullcreation}, \ref{halfcreation}, \ref{xtransf}, \ref{ptransf})
we find the transformation between oscillator modes:
\bea 
\; |a^e) &=& {1 \over \sqrt{2}} [ |a^L) + |a^R) ] 
\nonumber \\
\; |a^o) &=& {1 \over \sqrt{2}}\left[ C^+ [  |a^L) - |a^R) ] + C^- 
[ |a^{L\dagger}) - |a^{R\dagger}) ] \right]\nonumber\\
\; |a^L) &=& {1 \over \sqrt{2}} [|a^e) + C^{+T} |a^o) - C^{-T} 
|a^{o\dagger})]\nonumber\\
\; |a^R) &=& {1 \over \sqrt{2}} [|a^e) - C^{+T} |a^o) + C^{-T} |a^{o\dagger})]
\label{osctransf}
\eea
where the matrices $C^\pm $ are defined as
$$
C^\pm = \half [ E^{oo} A^T (E^{ee})^{-1} \pm (E^{oo})^{-1} A^T E^{ee} ]
$$
The half-string modes $a^L_{2n},\ a^{L\dagger}_{2n},\ a^R_{2n},
\ a^{R\dagger}_{2n}$
satisfy canonical commutation relations as can be seen from the
definition (\ref{halfcreation}) or from the properties of $C^\pm$:
\bea
C^{+T}C^{+} - C^{-T} C^{-} = 1&,&
C^{+}C^{+T} - C^{-} C^{-T} = 1\nonumber\\
C^{+T}C^{-} - C^{-T} C^{+} = 0 &,&
C^{+}C^{-T} - C^{-} C^{+T} = 0
\eea
\subsubsection{Half-string vacuum}
From (\ref{osctransf}) we see that, in terms of oscillators, the
 transformation to half-string 
modes is a Bogoliubov transformation under which creation and 
annihilation modes are mixed.  Hence the vacuum 
$|\O\rangle_L|\O\rangle_R$ annihilated by the half-string 
annihilation modes is not
the same as the original vacuum $|\O\rangle$ which is 
annihilated by the full-string 
annihilation modes.
From (\ref{osctransf}) it follows that they are related by:
\bea
|\O\rangle = \det\left(1 - (C^{+-1}C^-)^2 \right)^{26/4}
\exp - {1 \over 4} (a^{L\dagger} - a^{R\dagger}|C^{+-1}C^-
 |a^{L\dagger} - a^{R\dagger})|\O\rangle_L|\O\rangle_R\nonumber\\
|\O\rangle_L|\O\rangle_R =\det\left(1 - (C^{+T-1}C^{-T})^2 \right)^{26/4} 
\exp {1 \over 2}(a^{o\dagger}|C^{+T-1}C^{-T}
|a^{o\dagger})|\O\rangle
\label{vactransf1}
\eea
The components of the matrices  $C^{+-1}C^-$ and $ C^{+T-1}C^{-T}$
were already calculated in \cite{GJ1} where they enter as
Neumann coefficients in the definition of
the 4-point vertex. 
From definition  (2.40b) in \cite{GJ1}
of the matrix $V + \bar V$ we see that
$$
(V + \bar V)^{ee}= -2C^{+-1}C^-\qquad (V + \bar V)^{oo}= -2C^{+T-1}C^{-T}
$$
From equation (3.38a), the explicit matrix elements read:
\be
(C^{+T-1}C^{-T})_{2m+1, 2n+1} = - {\sqrt{(2m+ 1) ( 2n + 1)} \over 2m+ 1 + 2n +
  1} \left(  \begin{array}{c} - \half \\ m \end{array} \right)\left(  
\begin{array}{c} - \half \\ n \end{array} \right)
\label{coeffs}
\ee
It will also be useful to work out the transition between the vacua at 
zero momentum which we will denote by $|0\rangle_L|0\rangle_R$ and $|0\rangle$.
In terms of full  string modes, the state  $|0\rangle_L|0\rangle_R$
satisfies:
\bea
p_0|0\rangle_L|0\rangle_R&=& 0 \label{cond1}\\
\sum_{n=0}^\infty { (-1)^n \over 2n+ 1} ( a_{2n+1} +
a^\dagger_{2n+1})|0\rangle_L|0\rangle_R &=& 0\label{cond2}\\
a_{2m} |0\rangle_L|0\rangle_R &=& 0 \qquad m \neq 0\label{cond3}\\
\sum_{n=0}^\infty ( C^{+T}_{2m,2n+1 } a_{2n+1} - C^{-T}_{2m,2n+1 } 
a^\dagger_{2n+1})|0\rangle_L|0\rangle_R &=& 0 \qquad m \neq 0\label{cond4}
\eea
The state satisfying these conditions is given by:
\be
|0\rangle_L|0\rangle_R = \det\left(1 - (C^{+T-1}C^{-T})^2 \right)^{26/4}
\exp  {1 \over 2}(a^{o\dagger}|C^{+T-1}C^{-T}
|a^{o\dagger})|0\rangle.
\label{halfvac}
\ee
Conditions (\ref{cond1},\ref{cond3},\ref{cond4}) are trivially satisfied
so we only have to check  (\ref{cond3}). This condition is found to hold by
using the identity 
$$
\sum_{n=0}^\infty {1 \over 2n + 1 + a}{\Gamma(n+ \half) \over \Gamma(n + 1)} =
{\pi \over a} {\Gamma({a+1\over 2}) \over \Gamma({a\over2})}
$$ 
In \cite{GRSZ1, schnabl, GRSZ2}, 
the half-string vacuum $|0\rangle_L|0\rangle_R$ is also called 
the "butterfly state" and the expression
(6.42) in \cite{GRSZ2}, derived using conformal field theory methods, 
agrees with
(\ref{halfvac}).
\subsubsection{Half-string Neumann coefficients}
Using the relations (\ref{osctransf})
 we can express various quantities entering in the
definition of the string field theory action in terms of half-string 
oscillators. In principle, one could start from the full-string 
oscillator expressions
derived in \cite{GJ1} and perform the Bogoliubov transformation 
(\ref{osctransf})
to half string-oscillators but this is rather cumbersome. It is more 
convenient to start from the original position space
expressions in \cite{Witten} which are naturally expressed in terms of
half string modes.\\
The identity state in the matter sector $|\ci^X\rangle$ defined so that
$\langle \ci^X | \Psi \rangle = \int \Psi$, has the position space expression
$$
\ci^X[x^L, x^R] = \delta (x^L - x^R)
$$
where $\delta (x^L - x^R) \equiv \prod_{n=0}^\infty \delta( x^L_{2n} - x^R_{2n})$.
The oscillator expression of $|\ci^X\rangle$ can be found by writing
$$
|\ci^X\rangle = \int [Dx^L Dx^R] \ci[x^L, x^R] |x^L\rangle |x^R\rangle
$$
and using the oscillator expression for the 
half-string position eigenstates
\be
|\{x^L_{2n}\} \rangle = N^X_\half  \exp \left[ - \half
 (x^L | (E^{ee})^{ -2}| x^L) - 
\sqrt{2} i (a^{\dagger L}|
(E^{ee})^{-1}|x^L) + \half (a^{\dagger L}|a^{\dagger R})\right] |\O\rangle_L
\label{poseigenstate}
\ee
where  
$$
N^X_\half = \left({2 \over \pi}\right)^{26/4} \prod_{n=1}^\infty 
\left({2n \over \pi}
\right)^{26/4}
$$
(a similar expression holds for the right position eigenstates). 
Performing the Gaussian integral one finds
\be
|\ci^X \rangle = N_\ci 
\exp - (a^{L\dagger}|a^{R\dagger}) |\O\rangle_L|\O\rangle_R
\label{I}
\ee
with
$$
N_\ci = (N^X_\half)^2 \det (\pi (E^{ee})^2)^{26/2} = 1.
$$
The two-point vertex $|V_2^X\rangle_{12}$ is defined by $_{12}\langle V_2^X| |
\Phi \rangle_1 |\psi \rangle_2 = \int \Phi \star \Psi$. 
Its position space representation reads:
$$
V_2^X(x^L_1, x^R_1; x^L_2, x^R_2) = \delta (x^L_1 - x^R_2) \delta ( x^R_1 - 
x^L_2)
$$
Again using (\ref{poseigenstate}) 
we get 
\be
|V_2^X\rangle_{12} = \exp - \left[ ( a_1^{L\dagger } | a_2^{R\dagger} ) +(
a_2^{L\dagger } | a_1^{R\dagger})\right]
(|\O\rangle_L|\O\rangle_R)_{12}
\label{v2}
\ee
Equally simple is the expression for the three-point vertex 
$|V_3^X\rangle_{123}$
defined by \linebreak $\;_{123}\langle V_3^X| |
\Phi \rangle_1 |\Psi \rangle_2 |\X \rangle_3 = \int \Phi  \star \Psi \star \X$. 
From the position-space expression
$$
V_3^X(x^L_1, x^R_1; x^L_2, x^R_2;x^L_3, x^R_3)=
\delta (x^L_1 - x^R_3) \delta ( x^L_2 - x^R_1)\delta (x^L_3 - x^R_2)
$$
one gets
\be
|V_3^X\rangle_{123}= \exp - \left[ ( a_1^{L \dagger} | a_3^{R\dagger} ) +
( a_2^{L\dagger} | a_1^{R\dagger}) 
+( a_3^{L\dagger} | a_2^{R\dagger})\right]
(|\O\rangle_L|\O\rangle_R)_{123}.
\label{v3}
\ee
The state $|V_3^X\rangle_{123}$ can also be used to calculate star products 
in the oscillator representation. The star product of two real string fields
$|\Phi\rangle$ and $|\Psi\rangle$ is given by
\be
|\Phi \star \Psi \rangle_1 =\; _2 \langle \Phi |_3\langle \Psi| 
|V_3^X\rangle_{123}.
\label{star}
\ee
As a check on the normalization of $|V_3^X\rangle_{123}$ in (\ref{v3}), 
one can verify
that the rank one projector $|\O\rangle_L |\O\rangle_R$ indeed star-multiplies
to itself under (\ref{star}).

Comparing (\ref{v3}) with the  expression in terms of 
full-string modes derived 
in \cite{GJ1}, we see that the transformation to half-string modes has 
drastically simplified the Neumann matrices $V^{12}$ and $V^{21}$, while
 the matrix $V^{11}$ now even 
vanishes. 
\section{Some solutions to the projection equation}
\label{examples}
In this section we review some solutions to the projection equation
$\Psi = \Psi \star \Psi$ which have appeared in the literature.
We will restrict attention to states $|\Psi\rangle$ satisfying
\be
p^L_0  |\Psi\rangle = p^R_0 |\Psi\rangle = 0.
\label{cond}
\ee
Such states are invariant under separate translations of the 
left and right halves of the string.
The solutions satisfying this condition will turn out to have 
a direct counterpart in the ghost sector.
They 
can all be written as the vacuum 
$|0\rangle_L |0\rangle_R$ acted on with creation operators of strictly positive
mode number.
Note that the identity state $|\ci\rangle$ satisfies $(p_0^L + p_0^R)
 |\ci\rangle= 0$ 
but is not
an eigenstate of $p_0^L - p_0^R$, hence it does not belong to the class of
solutions we want to consider. 

Of particular importance in the construction of D-brane solutions in VSFT are
the  projectors of rank one \cite{RSZ3, GT1}. These are constructed from
any normalized state $|\chi\rangle_\half$ in  $\ch^\half$:
$$
\hat \Psi = |\chi \rangle_\half\;_\half \langle \chi |.
$$
Their position-space form is given by
$$ 
\Psi[ x^L, x^R ] = \chi[ x^L ] \chi^* [x^R]
$$
and the corresponding Fock space state is
$$
|\Psi\rangle = |\chi \rangle_L |\chi^* \rangle_R
$$
where $|\chi^*\rangle$ denotes the state with wavefunctional $\chi[x]^*$.
\subsection{Half-string vacuum}
The simplest example of a rank one projector 
is provided by taking $|\chi\rangle_\half = |0\rangle_\half$.
The corresponding solution is the half-string vacuum or butterfly
state $|B \rangle= |0\rangle_L|0\rangle_R$, 
whose full-string form was given in in
(\ref{halfvac}):
$$
|B \rangle= |0\rangle_L|0\rangle_R =   
\det\left(1 - (C^{+T-1}C^{-T})^2 \right)^{26/4}
\exp - {1 \over 2}(a^{o\dagger}|C^{+T-1}C^{-T}
|a^{o\dagger})|\O\rangle
$$
\subsection{D-25 brane sliver}
This solution, which describes a single D-25 brane, was found initially 
\cite{KP} in full-string variables where it takes the form of a squeezed state
$$
|\Sigma \rangle = \det (1 - S^2)^{26/4}
 \exp - \half \( a^\dagger| S |a^\dagger \) |0\rangle
$$
here we have introduced a new notation for sums that exclude 
the zero mode: $\( a | b \) \equiv \sum_{n=1}^\infty a_n b_n$.  
The matrix $S$ is related to the Neumann matrix $V^{11}$ \cite{GJ1} by:
\be
CS = { 1 \over 2 X} \left( 1 + X - \sqrt{(1 + 3 X)(1-X)} \right)
\label{Smatrix}
\ee
where $C$ is the twist matrix $C_{mn} = (-1)^n \delta_{mn}$ and
$X = C V^{11}$. 
The sliver state $|\Sigma \rangle$ was shown to satisfy (\ref{cond})
in \cite{MT}. Evidence for the fact that it corresponds to a rank 
one projector in half-string
space was found in \cite{RSZ3} and a proof was given in \cite{Moeller}. 
The corresponding half-string state $|\chi_\Sigma\rangle_\half$ 
is again a squeezed state:
$$
|\chi_\Sigma\rangle_\half \propto \exp - \half \( a^{\half\dagger} |
 {1 - D \over 1 + D} 
|a^{\half \dagger} \) |0\rangle_\half
$$ 
The D-25 brane sliver can be seen as the projector on the vacuum 
for a set of half-string oscillators related to the original ones by a 
Bogoliubov transformation.
The matrix $D$ can be expressed in terms of previously defined matrices
using (2.30-2.31) 
in \cite{Moeller}.
One finds
$$
D = \tilde E^{ee} \tilde A (E^{oo})^{-1}
\sqrt{E^{oo}\tilde A^T (\tilde E^{ee})^{-2}\tilde A E^{oo}}(E^{oo})^{-1}
\tilde A^T \tilde E^{ee}
$$ 
where a $\tilde{\ }$ denotes the submatrix obtained by excluding 
the zero mode.
One should be careful in trying to simplify this expression since
operators like $\sqrt{\tilde E^{ee} \tilde A (E^{oo})^{-1}}$ are ill-defined.
\subsection{GRSZ projectors}
In \cite{GRSZ2}, a class of rank one projectors, 
including the  D-25 brane sliver and the half-string vacuum
as special cases, 
 was constructed
using CFT methods. 
These arise from surface states in CFT and  they all satisfy the
condition (\ref{cond}). In terms of oscillators, all these projectors
 are squeezed states of the form
$$
|\X\rangle = \det (1 - V^2)^{26/4}
 \exp - \half \( a^\dagger | V |  a^\dagger\) |0\rangle.
$$
The condition (\ref{cond}) implies that the vector $v$ defined by
$$
v_{2n+1} = {(-1)^n \over \sqrt{2n+1}}; \qquad v_{2n} =0
$$
is an eigenvector of $V$ with eigenvalue 1.
The simplest example of this construction is provided by taking $V$ 
to be the identity matrix.
The corresponding projector $|\cn\rangle$ is called the {\em nothing state}
in \cite{GRSZ2}:
$$
|\cn\rangle = \exp - \half \( a^\dagger |  a^\dagger\) |0\rangle.
$$
The corresponding half-string state $|\chi_N \rangle_\half$ is given by:
$$
|\chi_\cn \rangle_\half=\exp 
 - \half \(a^{\half\dagger} | a^{\half\dagger}\) |0\rangle_\half.
$$
\section{Vacuum string field theory: ghost sector}\label{ghost}
\subsection{Bosonization conventions}
We will work in the bosonized formalism in which the $b,\ c$ ghosts
are be expressed in terms of a scalar linear dilaton field $\phi(\sigma)$
\cite{GSW, Witten}.
The mode expansion is
\begin{equation}
\phi (\sigma) = \phi_0 + {\sqrt{2} }\sum_{n = 1}^{ \infty}
\phi_n \cos (n \sigma)\,.
\end{equation}
and similarly for the conjugate momentum $\pi(\sigma)$: 
\begin{equation}
\pi (\sigma) ={1 \over \pi} \left( \pi_0 + {\sqrt{2} }\sum_{n = 1}^{ \infty}
\pi_n \cos (n \sigma)\right) \,.
\end{equation}
with
$$ [\phi_n,\pi_m] = i \delta_{mn}.$$
The momentum zero-mode $\pi_0$ plays the role of the ghost number
and is quantized in half-integer units.
The string field $\Psi$ entering in the VSFT action has ghost number
$\pi_0 = - \half$ while the string field parametrizing gauge transformations
has $\pi_0 = - {3 \over 2}$.
Creation and annihilation operators $d_n,\ d^*_n$ are defined by
\bea
|\phi) = {i \over \sqrt{2}} E [ |d) - |d^*) ]&\qquad&
|\pi) = {1 \over \sqrt{2}} E^{-1} [ |d) + |d^*) ]\nonumber\\
|d) = {1 \over \sqrt{2}} [ E |\pi) - i E^{-1} |\phi)]&\qquad&
|d^*) = {1 \over \sqrt{2}} [ E |\pi) + i E^{-1} |\phi)]
\label{fullcreationgh}
\eea
with commutation relations:
$$
[d_m, d_n^*]= \delta_{mn}.
$$
The  ghost fields $b_\pm,\ c^\pm$ with mode expansions
\bea
c^\pm (\sigma) &=& \sum_{n \in {\bf Z} } c_n e^{\pm i n \sigma}\\
b_\pm (\sigma) &=& \sum_{n \in {\bf Z} } b_n e^{\pm i n \sigma}
\eea
are bosonized according to:
\bea
c^+ (\sigma) & = & : e^{i \phi^+ (\sigma)}:
= e^{i \phi_0} e^{i \sigma (\pi_0 + 1/2)}
e^{\sum_{n = 1}^{ \infty} \frac{1}{\sqrt{n}} e^{in \sigma} d_{n}^* }
e^{-\sum_{m = 1}^{ \infty} \frac{1}{ \sqrt{m}} e^{-im \sigma} d_m }
\nonumber \\
c^- (\sigma) & = & : e^{i \phi^- (\sigma)}:
= e^{i \phi_0} e^{-i \sigma (\pi_0 + 1/2)}
e^{\sum_{n = 1}^{ \infty} \frac{1}{\sqrt{n}} e^{-in \sigma} d_{n}^* }
e^{-\sum_{m = 1}^{ \infty} \frac{1}{ \sqrt{m}} e^{im \sigma} d_m }\nonumber \\
b_+ (\sigma) & = & : e^{-i \phi^+ (\sigma)}:
= e^{-i \phi_0} e^{-i \sigma (\pi_0 - 1/2)}
e^{-\sum_{n = 1}^{ \infty} \frac{1}{\sqrt{n}} e^{in \sigma} d_{n}^* }
e^{\sum_{m = 1}^{ \infty} \frac{1}{ \sqrt{m}} e^{-im \sigma} d_m }
\nonumber \\
b_- (\sigma) & = & : e^{-i \phi^- (\sigma)}:
= e^{-i \phi_0} e^{i \sigma (\pi_0 - 1/2)}
e^{-\sum_{n = 1}^{ \infty} \frac{1}{\sqrt{n}} e^{-in \sigma} d_{n}^* }
e^{\sum_{m = 1}^{ \infty} \frac{1}{ \sqrt{m}} e^{im \sigma} d_m }
\label{bosformula}
\eea
Hermitean conjugation works on the ghost modes as $c_n^\dagger = c_{-n}$,  
so the ghost fields  $c^\pm (\sigma), b_\pm(\sigma)$ 
are Hermitean. From (\ref{bosformula}) it follows that  
Hermitean conjugation acts
on the bosonized ghosts modes as: 
\be
d_n^\dagger = - d_n^*; \qquad
\phi_n^\dagger = - \phi_n; \qquad
\pi_n^\dagger = - \pi_n
\label{hc}
\ee
This leads to slight differences with the matter sector so
we proceed to give some further conventions in the ghost sector.
As before, we define a vacuum $|\O\rangle$
satisfying $d_n |\O\rangle = 0$, and its Hermitean conjugate
$\langle \O | = (|\O\rangle)^\dagger$ satisfying $\langle \O |d^*_n =0$
and \linebreak $\langle \O | \O \rangle = 1$. Position eigenstates have the 
oscillator
expression
\be
|\phi\rangle \equiv |\{\phi_n\} \rangle = N^\phi \exp \left[ - 
\half (\phi | E^{ -2}| \phi) - 
\sqrt{2} i (d^*|
E^{-1}|\phi) + \half (d^*|d^*)\right] |\O\rangle,
\ee
with $N^\phi = (N^X)^{1/26}$, while their Hermitean conjugates read
\be
\langle\phi|\equiv (|\phi \rangle)^\dagger 
 = N^\phi \langle \O | \exp \left[ - \half (\phi | E^{ -2}| \phi) - 
\sqrt{2} i (d|
E^{-1}|\phi) + \half (d|d)\right].
\ee
These satisfy
$$
\langle \phi | \hat \phi_n = \langle \phi | (- \phi_n)
$$
where $\hat \phi_n$ refers to the operator and $\phi_n$ to 
its eigenvalue.
The inner product reads
$$
\langle \phi ' | \phi \rangle = \delta (\phi + \phi')
$$
Hence one finds the completeness relation
$$
{\bf 1} = \int [D\phi] |\phi\rangle \langle -\phi|
$$
Similarly, for momentum eigenstates one has
$$
\langle \pi | \hat \pi_n = \langle \pi | (- \pi_n); \qquad
\langle \pi ' | \pi \rangle = \delta (\pi + \pi').
$$
One goes to the position-space representation by expanding
$$
|\Psi \rangle = \int [D \phi] \Psi[ \phi ] | \phi \rangle
$$
where
$$
\Psi[ \phi ] = \langle - \phi | \Psi \rangle.
$$
The inner product can be written as
$$
\langle \Phi | \Psi \rangle =  \int [D \phi] \Phi^*[-\phi] \Psi[ \phi].
$$

\subsection{Half-string modes}
The formulae for the transition to half-string modes derived for
the matter sector go through for the ghost sector as well
under the substitutions:
$$
x \to \phi \ \ \ p \to \pi\ \ \  a \to d, \ \ \  a^\dagger \to d^*.
$$
Half-string modes are defined by
\bea
|\phi^L) = {1 \over \sqrt{2} } [ |\phi^e) + A |\phi^o) ]&\ &
|\phi^R) = {1 \over \sqrt{2} } [ |\phi^e) - A |\phi^o) ]\nonumber\\
|\pi^L) = {1 \over \sqrt{2} } [ |\pi^e) + A |\pi^o) ]&\ &
|\pi^R) = {1 \over \sqrt{2} } [ |\pi^e) - A |\pi^o) ]
\label{halfmodesgh}
\eea
While the transformation between oscillator modes reads:
\bea 
\; |d^e) &=& {1 \over \sqrt{2}} [ |d^L) + |d^R) ] 
\nonumber \\
\; |d^o) &=& {1 \over \sqrt{2}}\left[ C^+ [  |d^L) - |d^R) ] + C^- 
[ |d^{L\dagger}) - |d^{R\dagger}) ] \right]\nonumber\\
\; |d^L) &=& {1 \over \sqrt{2}} [|d^e) + C^{+T} |d^o) - C^{-T} 
|d^{o\dagger})]\nonumber\\
\; |d^R) &=& {1 \over \sqrt{2}} [|d^e) - C^{+T} |d^o) + C^{-T} |d^{o\dagger})]
\label{osctransfgh}
\eea
Full- and half-string vacua are related through
$$
|\O\rangle_L|\O\rangle_R = \det\left(1 - (C^{+T-1}C^{-T})^2 \right)^{1/4} 
\exp - {1 \over 2}(d^{o*}|C^{+T-1}C^{-T}
|d^{o*})|\O\rangle
\label{vactransfgh}
$$
\subsection{Vertices and midpoint insertions}
The states defining the one- and three-point vertices in the ghost sector
differ from the
pure delta-function overlaps of the matter sector through the
presence of midpoint insertions. These have their origin in the
fact that these vertices introduce delta-function curvature
 on the string world-sheet to which the linear dilaton couples. 
Let us first introduce  states $|\ci^\phi\rangle,\ |V_2^\phi\rangle_{12}$ and 
$|V_3^\phi\rangle_{123}$ defining pure delta-function overlaps analogous
to the ones used in the matter sector:
\bea
\ci^\phi (\phi^L, \phi^R)&=& \delta (\phi^L_1 - \phi^R_2)\\
V_2^\phi (\phi^L_1, \phi^R_1; \phi^L_2, \phi^R_2) &=&
\delta (\phi^L_1 - \phi^R_2) \delta ( \phi^R_1 - \phi^L_2)\\
V_3^\phi (\phi^L_1, \phi^R_1; \phi^L_2, \phi^R_3;\phi^L_2, \phi^R_3)&=&
\delta (\phi^L_1 - \phi^R_3) \delta ( \phi^L_2 - \phi^R_1)
\delta ( \phi^L_3 - \phi^R_2).
\eea
In terms of oscillators, these are given by (cfr. (\ref{I}, \ref{v2}, 
\ref{v3}))
\bea
|\ci^\phi \rangle &=& \exp - (d^{L*}|d^{R*}) |\O\rangle_L|\O\rangle_R\\
|V_2^\phi\rangle_{12} &=& \exp - [ ( d_1^{*L} | d_2^{*R} ) +( d_2^{*L} | d_1^{*R})]
(|\O\rangle_L|\O\rangle_R)_{12}\label{v2gh}\\
|V_3^\phi\rangle_{123} &=& \exp - [ ( d_1^{*L} | d_3^{*R} ) +( d_2^{*L} | d_1^{*R}) 
+( d_3^{*L} | d_2^{*R})]
(|\O\rangle_L|\O\rangle_R)_{123}
\eea
The ghost sector vertices $|\ci^{gh},\ \rangle |V_2^{gh}\rangle_{12}$ and $
|V_3^{gh}\rangle_{123}$ include midpoint insertions:
\bea
|\ci^{gh} \rangle &=& e^{-{3 \over 2}i \phi(\pi/2)} |\ci^\phi \rangle\\
|V_2^{gh}\rangle_{12} &=&|V_2^\phi\rangle_{12}\\
|V_3^{gh}\rangle_{123}&=& e^{{i \over 2}( \phi_1(\pi/2)+  \phi_2(\pi/2)+ 
\phi_3(\pi/2))}|V_3^\phi\rangle_{123}.
\label{ghostvertices}
\eea
In the last expression, we have to chosen to divide the midpoint 
insertion evenly over the three copies of the ghost Fock space 
labeled by $_1,\ _2,\ _3$. This is possible because of the
delta-function character of $V_3^\phi$.
Note that, by making a field redefinition, one could absorb the
midpoint insertions of the three-point vertex in the definition 
of the string field, but this would introduce a 
midpoint insertion in the two-point vertex.
Hence it seems hard to get rid of the midpoint insertions in the string field 
theory action. However, we will see that if the quadratic term in the action
contains the BRST operator proposed in \cite{GRSZ1}, which is itself a midpoint
insertion, it will be possible to effectively get rid of all the
midpoint insertions in the action through a field redefinition. 
\subsection{BRST operator}
In \cite{GRSZ1}, it was proposed that the BRST operator 
governing the VSFT action is a pure midpoint insertion: 
$$
Q = {1 \over 2} [ c^+(\pi/2) + c^- (\pi/2)].
$$
Using (\ref{bosformula}) and the Baker-Campbell-Hausdorff formula, we 
can rewrite $Q$ in the bosonized form
\be
Q = {\kappa_1 \over 2} e^{i \phi(\pi/2)}\left[e^{i {\pi \over 2} (\pi_0 + \half)} 
e^{ i \sqrt{2} \sum_{n=0}^\infty {(-1)^n \over 2n+1} \pi_{2n+1} } 
+e^{- i {\pi \over 2} (\pi_0 + \half)} 
e^{- i \sqrt{2} \sum_{n=0}^\infty {(-1)^n \over 2n+1} \pi_{2n+1} }\right]
\label{midQ}
\ee
where $\kappa_1$ is an (infinite) constant: 
$\kappa_1 = \exp \sum_{n=1}{1 \over 2n}$.
This constant can be absorbed into a redefinition of the string 
field and the overall normalization of the action. 
By definition, the states entering the string field
theory action have fixed ghost number
$\pi_0 = - \half$. The factors $e^{\pm i {\pi \over 2} (\pi_0 + \half)}$
 in (\ref{midQ}) act trivially on such states. 
The remaining dependence of $Q$ on 
the momentum modes becomes very simple in terms of half-string modes:
using (\ref{halfmodesgh}, \ref{Adef}) we have
$$
\sqrt{2} \sum_{n=0}^\infty {(-1)^n \over 2n+1} \pi_{2n+1} = {\pi \over 2 
\sqrt{2}} (\pi_0^L - \pi_0^R)
$$
Hence, when acting on states with $\pi_0 = - \half$,  $Q$ is equivalent to
the operator $\kappa_1 e^{i \phi(\pi/2)} Q'$ where, for later convenience, we have
defined
\be
Q'=  \cos {\pi \over 2 \sqrt{2}} (\pi_0^L - \pi_0^R).
\label{Q'def}
\ee

\section{Ghost sector projection equation}\label{redef}
\subsection{Field redefinition}
In this section we propose a field redefinition which considerably
simplifies the equation of motion in the ghost sector of VSFT.
This equation arises from the action
\be
S^{gh}[\Psi] = - \kappa_0 \left[ \half \;_{12}\langle V_2^{gh}| |\Psi\rangle_1
Q|\Psi\rangle_2 + {1 \over 3} \;_{123}\langle V_3| |\Psi\rangle_1|\Psi\rangle_2
|\Psi\rangle_3 \right]
\label{ghaction}
\ee
with $\kappa_0$ an overall normalization constant.
Using the results of the previous section, the quadratic term
can be rewritten as
$$
\half\,_{12}\langle V_2^{gh} | |\Psi\rangle_1  Q |\Psi\rangle_2
= {\kappa_1 \over 2} \,_{12}\langle V_2^\phi ||\Psi\rangle_1 
e^{i \phi(\pi/2)} Q' |\Psi\rangle_2.
$$
We can use the fact that $|V_2^\phi\rangle$ satisfies the overlap equations
\be
\ (\phi_{1,2n} - \phi_{2,2n}) |V_2^\phi\rangle_{12} = 0
\ee
(this can be derived, e.g., from the oscillator expression (\ref{v2gh})),
to evenly distribute the $e^{i \phi(\pi/2)}$ insertion
 over the two copies of the ghost
Fock space labeled by $_1$ and $_2$:  
\bea
\half\,_{12}\langle V_2^{gh} | |\Psi\rangle_1  Q |\Psi\rangle_2=
{\kappa_1 \over 2}\,_{12}\langle V_2 | 
e^{{i \over 2}\phi(\pi/2)}|\Psi\rangle_1  
e^{{i \over 2}\phi(\pi/2)} Q' 
|\Psi\rangle_2 \nonumber\\
={\kappa_1 \over 2}\,_{12}\langle V_2 | e^{{i \over 2}\phi(\pi/2)} 
|\Psi\rangle_1 
 Q' e^{{i \over 2}\phi(\pi/2)}
|\Psi\rangle_2
\eea
where, in the second line, we used the fact that $[ \phi(\pi/2),(\pi_0^L - \pi_0^R)]
=0$.
Using the definition (\ref{ghostvertices})
 of the three-point vertex, the action (\ref{ghaction})
 reduces to
\bea
S^{gh}[\Psi] &=& - {\kappa_0}\Big[ 
{\kappa_1 \over 2}\,_{12}\langle V_2^\phi | e^{{i \over 2}\phi(\pi/2)}
|\Psi\rangle_1  
 Q' 
e^{{i \over 2}\phi(\pi/2)} | \Psi\rangle_2\nonumber\\ 
&+& {1 \over 3}\,_{123}\langle 
V_3^\phi | e^{{i \over 2}\phi(\pi/2)}| \Psi\rangle_1
e^{{i \over 2}\phi(\pi/2)}|\Psi\rangle_2 e^{{i \over 2}\phi(\pi/2)}
|\Psi\rangle_3 \Big]
\eea
We can now absorb the midpoint insertions $e^{{i \over 2}\phi(\pi/2)}$ into 
 a field redefinition 
\be
| \Psi' \rangle 
= -{1 \over  \kappa_1} e^{{i \over 2}\phi(\pi/2)}
|\Psi \rangle.
\label{fieldredef}
\ee
We get
$$
S^{gh}[\Psi'] = -{\kappa_0'} \left[ \half\,_{12}\langle V_2^\phi | 
\left(|\Psi'\rangle_1  
Q'
|\Psi'\rangle_2\right) - {1 \over 3}\,_{123}\langle V_3^\phi | 
\left(|\Psi'\rangle_1
|\Psi'\rangle_2|\Psi'\rangle_3\right) \right]
$$
where $ \kappa_0' = \kappa_0 (\kappa_1)^3$.
The equation of motion becomes:
\be
Q' |\Psi'\rangle 
=|\Psi'\rangle  \star'
|\Psi'\rangle
\label{neweom}
\ee
where $\star'$ is determined by the vertex 
$|V_3^\phi \rangle$ without midpoint
insertions and has the
same form as the star product (\ref{star}) in the matter sector.

Since $|\Psi\rangle$ was restricted to have ghost number $-\half$,
the new string field $|\Psi'\rangle$ is constrained
to have ghost number zero:
\be
(\pi_0^L + \pi_0^R)|\Psi'\rangle = 0.
\label{totgh}
\ee
When we restrict attention to states which in addition are 
eigenstates of $\pi_0^L$ and $\pi_0^R$ separately, i.e. states satisfying
\be
\pi_0^L |\Psi'\rangle = -\pi_0^R |\Psi'\rangle = \lambda 
|\Psi'\rangle
\label{constr}
\ee
for some $\lambda$, the action of the operator $Q'$
 in (\ref{neweom}) reduces to multiplication by
a constant.
Note that, although the full-string ghost number $\pi_0$ was quantized,
there is no a priori restriction on the value of the 
half-string ghost number $\lambda$.  We should, however, make sure that the
restriction (\ref{constr}) is compatible with the reality condition on 
the string field, which in the ghost sector reads \cite{Witten}
\be
\Psi'[\phi^L, \phi^R] = \Psi'^*[-\phi^R,-\phi^L].  
\label{realgh}
\ee
This restricts the allowed values of $\lambda$ to $\lambda=0$. Indeed, 
for general $\lambda$, the zero-mode part of $\Psi'[\phi^L, \phi^R]$ is
$e^{i \lambda( \phi_0^L - \phi_0^R)}$. Hence the reality condition
is satisfied only for $\lambda =0$.

In summary, a class of solutions to
 the equations of motion in the ghost sector
of VSFT is provided by states $|\Psi'\rangle$
that satisfy the condition
\be
\pi_0^L |\Psi'\rangle = \pi_0^R |\Psi'\rangle = 0
\label{condgh}
\ee
and  are solutions of the projection equation
\be
|\Psi'\rangle =|\Psi'\rangle \star' |\Psi'\rangle .
\label{proj}
\ee
In addition, they should satisfy the reality condition 
(\ref{realgh}). 

The condition (\ref{condgh}) can be interpreted as stating that 
the ghost number of $|\Psi'\rangle$ (and hence the ghost number
 of the original field $|\Psi\rangle$ as well)
should be evenly
distributed over the left- and right halves of the string.

\subsection{Operator formulation}\label{operator}
The absence of midpoint insertions in $\star'$ implies that we can write 
(\ref{proj}) in the form of a simple algebraic equation
for operators in half-string space, just as in the matter case.
 The map between string fields $|\Psi'\rangle$ 
and half-string operators $\hat \Psi'$ is
slightly  different from the one in the matter sector because $\phi$ is
antihermitean. It reads:
\bea 
|\Psi' \rangle &=& \int [D\phi^L D\phi^R]\  \Psi' [\phi^L , \phi^R] 
|\phi^L\rangle 
 |\phi^R\rangle \\
& \Updownarrow & \\
\hat \Psi' &=& \int[D\phi^L D\phi^R] \  \Psi' [\phi^L, \phi^R] |\phi^L\rangle 
\langle- \phi^R |
\eea 
With this definition, $\star'$ multiplication reduces to
ordinary multiplication of operators:
$$
\Phi' \star' \Psi' \Leftrightarrow \hat \Phi' \hP'.
$$
The reality condition on the string field (\ref{realgh}) again reduces to
a Hermiticity condition on operators:
$$
\Psi'[\phi^L, \phi^R] = \Psi'^*[-\phi^R,-\phi^L]\Leftrightarrow 
\hat {\Psi}' = \hat {\Psi}'^\dagger.
$$
The operator form of the conditions
 (\ref{realgh}, \ref{condgh}) and the
equation of motion (\ref{proj}) is
\bea
\;  \hat {\Psi}' &=&  \hat {\Psi}'^\dagger \label{realop}\\
\; \pi_0^\half\ \hat {\Psi}'&=&\hat {\Psi}'\ \pi_0^\half =0 \label{gncond}\\
\;\hat {\Psi}' &=& \hat {\Psi}'^2 \label{projop}
\eea
 The ghost number condition (\ref{gncond}), together
with Hermiticity (\ref{realop}), restricts the zero mode part of $\hat \Psi'$
to be $|\pi_0 =0 \rangle_\half \;_\half\langle \pi_0 =0|$. Hence we can 
decompose
\be
\hat {\Psi'} = \left(|\pi_0 =0 \rangle_\half \;_\half\langle \pi_0 =0|\right)
\otimes \hat M.
\label{zerofix}
\ee
where $\hat M$ works on the subspace orthogonal to the zero mode.
The equations (\ref{realop}, \ref{projop}) simply state that $\hat M$ 
is a Hermitean projection operator.

With the ghost zero-mode part constrained as in (\ref{zerofix}), 
VSFT becomes, at least formally, equivalent to a cubic $U(\infty)$  
matrix model. Indeed, if we define the operator $\hat P$ 
as the tensor product of $\hat M$
with the matter string field $\Psi^m$, $\hat P = \hat \Psi^m \otimes \hat M$, 
the total matter and ghost action becomes
$$
S[\hat P] = -\kappa_0' \Tr \left[ \half \hat P^2 - {1 \over 3} \hat P^3 \right]
$$
 which is invariant under the unitary transformations $\hat P \to
U \hat P U^\dagger$. Here, $U$ works on the half-string subspace 
orthogonal to the ghost zero mode. A similar $U(\infty)$ invariance 
is present in purely cubic string field theory \cite{horowitz}.
\subsection{Examples}
It is now a simple matter to construct ghost sector solutions.
 Using the decomposition (\ref{zerofix})
any Hermitean projection operator 
$\hat M$ in the subspace
of $\ch^\half$ 
not involving the zero mode
will give rise to a ghost sector solution.
In section \ref{examples}, we gave several examples of rank one 
projectors in the matter sector with  half-string momenta equal to zero. These
can all be mapped to projectors in the ghost sector satisfying
(\ref{gncond}, \ref{projop}). Their 
wavefunctionals factorize into a product of half-string Gaussians
so they also obey the reality condition (\ref{realgh}).

A first example is the {\em ghost butterfly state}
$$
|B'\rangle = 
|0\rangle_L |0\rangle_R =\det\left(1 - (C^{+T-1}C^{-T})^2 \right)^{1/4}
\exp - {1 \over 2}(d^{o*}|C^{+T-1}C^{-T}
|d^{o*})|0\rangle.
$$
The matrix elements of $C^{+T-1}C^{-T}$ were given in (\ref{coeffs}).

From the D-25 brane sliver in the matter sector one can construct an analogous
{\em ghost sliver state} given by
$$
|\Sigma'\rangle =  \det (1 - S^2)^{1/4}\exp - \half \(d^*| S |d^* \) |0\rangle.
$$
with the matrix $S$ given by (\ref{Smatrix}).

All projectors arising from surface states constructed in \cite{GRSZ2}
have analogous solutions in the ghost sector. The simplest example
is the {\em ghost sector nothing state} given by
$$
|\cn'\rangle = \exp - \half \(d^* | d^*\) |0\rangle.
$$

Except for the ghost sector nothing state, which is not normalizable, the 
solutions discussed here are rank one projectors for which the ghost
part of the action takes the value $- \kappa_0' / 6$. 
If we combine such a rank one ghost projector with a rank one matter solution,
 the total action will still be equal to $- \kappa_0' / 6$. This suggests the
identification
$$
\kappa_0'/6 = \ct_{25}
$$
where $\ct_{25}$ is the tension of the D-25 brane.

The fact that we find finite action solutions for finite values
of $\kappa_0'$
seems in contradiction with \cite{GRSZ1}, where it was found that
the analogous constant $\kappa_0^{GRSZ}$ should diverge in order to have
finite action solutions. However, one has 
to bear in mind that the normalization  
of the vertices in (\ref{ghostvertices}) in the
string field theory action differs from the one used in \cite{GRSZ1}
by singular factors \cite{GT2, Moeller}.
 A more precise derivation of the 
relation between $\kappa_0'$ and $\kappa_0^{GRSZ}$ would require
 a careful regularization
of such factors and is beyond the scope of the present work.
\section{Discussion}

In this paper we have focused on the ghost
sector of VSFT with the  BRST-operator $Q$ taken to be the 
pure midpoint insertion proposed in \cite{GRSZ1}. We have shown that, 
for this particular choice of $Q$, the main obstacles to 
constructing ghost sector solutions, viz. the presence of $Q$ in the
equations of motion and the midpoint insertion in the star product, can
be dealt with simultaneously by making a field redefinition. 
This redefinition removes the midpoint insertion from the star product and
the remaing kinetic operator $Q'$ depends, 
when expressed
in half-string modes, only on
the difference of the left- and right ghost numbers. 
Hence it can  be rendered 
trivial by consistently truncating the string field to have its ghost
number symmetrically divided over the left- and right halves of the string. 
The resulting equation of motion is a pure projection equation and solutions
can be found by using the techniques developed for the matter sector. 
In the light of certain subtleties in the transition to 
half-string variables, hinted at in section \ref{half} and discussed
more extensively in \cite{MT, bars, douglas}, 
we should note that the use of half-string variables is not
essential in our construction: it merely provides a simple interpretation
of the operator $Q'$ in terms of half string zero modes and 
facilitates the construction of solutions as projection operators on
half-string Hilbert space. We end with some observations and open
problems.
\begin{itemize}
\item It seems that the relation between the ghost sector equation 
of motion and the projection equation is a feature special to the
pure midpoint BRST-operator $Q$ in contrast to the other
pure ghost operators considered previously \cite{RSZ1}. Indeed, it is unlikely
that similar simplifications would go through with any of the other
pure ghost BRST operators. It is quite remarkable
that precisely this $Q$ has emerged as a likely candidate for the correct 
description of the ghost sector. It has been remarked in \cite{GRSZ1} that 
non-pure ghost BRST operators, obtained by tensoring $Q$ with a suitable 
matter part, could also be considered good candidates for 
describing VSFT. 
A generalization along these lines appears, in fact, to be necessary in order
to obtain the correct structure of gauge transformations in VSFT
\cite{Hashimoto:2002sm}.
The simplifications in the ghost sector discussed in this paper
would also go through for these operators.
\item
In this paper, we used the bosonized form of the ghost system.
It remains to be seen  whether a similar approach would work
in terms of fermionic ghosts  since the field redefinition (\ref{fieldredef})
has no local expression in terms of    $(b,\ c)$ fields.
\item
In \cite{GRSZ1}, solutions in the ghost sector of VSFT were constructed by
introducing new `twisted ghost' variables, in terms of which
these solutions reduced to pure projectors. 
The transition to twisted variables acts on the Virasoro generators as
$$ 
L_n^{tw} = L_n + n j_n+ \delta_{n,0}
$$
with $j_n$ the modes of the ghost number current. 
One might wonder how this approach  is
related to 
the one used in  the present paper. A first guess would
be that our field redefinition (\ref{fieldredef}) implements 
the transformation to the twisted ghost system:
$$
L_n^{tw} =\!\!\!\!\!\!{^{\;?}} L_n' \equiv e^{{i \over 2} \phi(\pi/2)} 
L_n e^{-{i \over 2}
 \phi(\pi/2)}
$$
 This is not the case
however as one can easily verify. It seems likely that both approaches are 
related but it is not yet clear how. 
\item
In section \ref{operator}, 
we found that, once one imposes the constraint (\ref{condgh}) on the
half-string ghost number,  
 VSFT can be written as a $U(\infty)$ matrix model. 
Such a $U(\infty)$ symmetry was also also argued to be present in the 
tachyonic vacuum from the low-energy point of view in the presence
of a large $B$-field \cite{Gopakumar:2001rw}.
It would be interesting to see how both points of view are related.
\end{itemize}
 
\acknowledgments
We are grateful to  D.\ Ghoshal, T.\ Jayaraman and V.\ A.\ Kosteleck\'y
for stimulating discussions.
J.R.\ gratefully acknowledges travel support from K.U.Leuven. J.R.\ also wishes
to thank the University of the Algarve, the Strings Workshop at IMSc Chennai
and the Harish-Chandra Research Institute
for hospitality.
R.P.\ acknowledges financial support from the Portuguese Funda\c c\~ao para
a Ci\^encia e a Tecnologia under grant number CERN/P/FIS/40108/2000.

\end{document}